\documentclass[manuscript]{aastex}
\usepackage{natbib}
\usepackage{color}
\usepackage{tabularx}

\usepackage{longtable}
\usepackage{multirow}

\slugcomment{ApJL, 851, 2.}
\shorttitle{Constraint the kilonova rate with DLT40 and the light curve of AT2017gfo/DLT17ck}
\shortauthors{Yang et al.}

\begin{document}
 \title{An empirical limit on the kilonova rate from the DLT40 one day cadence supernova survey}

\author{Sheng Yang,$\!$\altaffilmark{1,2} 
Stefano Valenti,$\!$\altaffilmark{1} Enrico Cappellaro,$\!$\altaffilmark{2}
David, J. Sand,$\!$\altaffilmark{3}  
\and
 Leonardo Tartaglia,$\!$\altaffilmark{3,1}  Alessandra Corsi,$\!$\altaffilmark{4} Daniel E. Reichart,$\!$\altaffilmark{5}
 \and
 Joshua Haislip,$\!$\altaffilmark{5} Vladimir Kouprianov$\!$\altaffilmark{5}
}

\begin{abstract}
Binary neutron star mergers are important to understand stellar evolution, the chemical enrichment of the universe via the r-process, the physics of short gamma-ray bursts, gravitational waves and pulsars. The rates at which these coalescences happen is uncertain, but it can be constrained in different ways. One of those is to search for the optical transients produced at the moment of the merging, called a kilonova, in ongoing SN searches. However, until now, only theoretical models for kilonovae light curve were available to estimate  their rates.%have been computed without knowing {\bf true kilonova light curve properties}. 
The recent kilonova discovery AT~2017gfo/DLT17ck gives us the opportunity to constrain the rate of kilonovae using the light curve of a real event. We constrain the rate of binary neutron star mergers using the DLT40 Supernova search, and the native AT~2017gfo/DLT17ck light curve obtained with the same telescope and software system. Excluding AT~2017gfo/DLT17ck due to visibility issues, which was only discovered thanks to the aLIGO/aVirgo trigger,%{\bf (due to its position relatively close to the Sun at the time of the merger)}, 
no other similar transients %transients similar to AT~2017gfo/DLT17ck have been 
detected during 13 months of daily cadence observations of $\sim$ 2200 nearby ($<$40 Mpc) galaxies. We find that the rate of BNS mergers is lower than  0.47 - 0.55 kilonovae per 100 years per $10^{10}$ $L_{B_{\odot}}$ (depending on the adopted extinction distribution). In volume, this translates to  $<0.99\times 10^{-4}\,_{-0.15}^{+0.19},\rm{Mpc^{-3}}\,\rm{yr^{-1}}$(SNe Ia-like extinction distribution), consistent with previous BNS coalescence rates.  Based on our rate limit, and the sensitivity of aLIGO/aVirgo during O2, it is very unlikely that kilonova events are lurking in old pointed galaxy SN search datasets.

\end{abstract}

 
\altaffiltext{1}{Department of Physics, University of California, 1 Shields Avenue, Davis, CA 95616-5270, USA}
\altaffiltext{2}{INAF Osservatorio Astronomico di Padova, Vicolo dell’Osservatorio 5, I-35122 Padova, Italy}
\altaffiltext{3}{Department of Astronomy/Steward Observatory, 933 North Cherry Avenue, Room N204, Tucson, AZ 85721-0065, USA}
\altaffiltext{4}{Physics $\&$ Astronomy Department, Texas Tech University, Lubbock, TX 79409, USA 0000-0003-3433-1492}
%\altaffiltext{5}{Department  of  Physics  and  Astronomy,  Rutgers,  TheState University of New Jersey, Piscataway, NJ 08854, USA}
\altaffiltext{5}{Department of Physics and Astronomy, University of North Carolina at Chapel Hill, Chapel Hill, NC 27599, USA}

\section{Introduction}
%Since the discovery of the first binary neutron star (BNS) system \citep{1975ApJ...195L..51H}, these events have been studied with great interest by the astronomical community because they provide fundamental information on the last stage of evolution of stars in binary system, they are the best cases where a precise measurement of the neutron star mass can be derived and one of possible sources of gravitational waves, detectable by aLIGO and aVirgo. 
Binary neutron star (BNS) systems \citep{1975ApJ...195L..51H} have been studied with great interest by the astronomical community because %they provide fundamental information on the last stage of evolution of stars in binary systems. They are the best cases where a precise measurement of the neutron star mass can be derived and one of possible sources of gravitational waves detectable by aLIGO and aVirgo.} 
of their connection with many open problem of astrophysics, from short GRB to r-process element production, from the physics of very high density matter to gravitational waves.
The number of known BNS today is limited to a dozen systems \citep{2012ARNPS..62..485L} and the rate 
of BNS coalescences is known with order of magnitudes of uncertainty\citep{2012MNRAS.425.2668C, 2013ApJ...767..140P, 2014MNRAS.437..649S, 2015ApJ...815..102F, 2015ApJ...814...58D, 2015ApJ...806..263D, 2016MNRAS.455...17V, 2015MNRAS.448..928K, 2010CQGra..27q3001A, 2017PRL119161101, 2015ApJ...811L..22J,2014ApJ...794...23D, 2013ApJ...779...18B}. 
The rate of BNS coalescences  can be constrained from the  population of galactic pulsars \citep{2004ApJ...601L.179K}, from modeling the evolution of binary system \citep{2015ApJ...814...58D, 2015ApJ...806..263D}, from the cosmic abundance of r-process elements \citep[][]{2016MNRAS.455...17V} or measuring the rate of short Gamma-Ray Bursts (GRBs), most likely produced at the moment of the coalescence \citep{2013ApJ...779...18B}.  

An alternative method to constrain the rate of BNS mergers is to constrain the rate of kilonovae detected in supernova (SN) search surveys.  Kilonovae are thought to be the ubiquitous, isotropically emitting %optical/NIR 
counterparts to neutron star mergers. They are expected to eject at very high velocity a small mass (0.01-0.05 $M_{sun}$) mainly made of high opacity r-process heavy elements, %, and because they are powered by the r-process decay of heavy nuclei, they 
hence are predicted to have a `red' spectrum, to be faint at maximum light ($M_{V}\sim-16\,\rm{mag}$) and %to  that evolves 
declining quickly over the course of 1-2 weeks\citep[e.g.][]{2010MNRAS.406.2650M, 2013MNRAS.430.2121P, 2013ApJ...774...25K}.  This is in contrast to the most common SNe, which evolve %on $\tau\sim 10-100\,\rm{day}$
of $0-100\,\rm{day}$ timescales \citep[e.g. see Figure 1.1 of][]{Kasliwal11}.

One clear hindrance to calculate the kilonova rate has been the lack of observed kilonova events \citep[with possible exceptions, see][for a compilation]{Jin16} in ongoing transients searches.
%, which are necessary to calculate the 
%{\color{red} control time \footnote{\color{red} control time can be considered as the effective searching time, which is first defined by \cite{1942ApJ....96...28Z}, see section 3 for details.}}  for kilonovae in a given time domain search. 
Some progress has been made by using theoretical kilonova light curves %and using these 
to calculate %their apparent 
the upper limit of their rate in programs like the Dark Energy Survey \citep{2017ApJ...837...57D}. Additionally, there have been several recent attempts to estimate the rate of fast optical transients that evolve on sub-day timescales \citep[$\tau\sim0.5\,\rm{hours}$ to 1 day;][]{2013ApJ...779...18B} all the way up to $\sim10\,\rm{day}$ timescales \citep{2014ApJ...794...23D}.

With the discovery of AT~2017gfo \citep{2017PRL119161101,2017ApjMMA}, %and its high cadenced photometric follow-up, 
we can directly constrain the rate of kilonovae by using its light curve as a template%, along with some modifications to account
while allowing for possible diversity in kilonova light curves and a range of extinction values.  Here we present the %first 
rate estimate for kilonovae using our observed light curve of AT~2017gfo and the data from the ongoing Distance less then 40 Mpc (DLT40) Supernova search \citep{2017arXiv171103940T}.  
%
%\footnote{{\color{red} DLT40 is designed as a one day cadence supernovae search while the typically cadence of detected supernovae was less than 24 hours. For sure, there exists several supernova with more than 24 hours letance due to weather loss but the mainly detections could fulfill our demandings.}}
%
DLT40 is a SN search that points galaxies within $D\lesssim40\,\rm{Mpc}$ with a one day cadence \footnote{see next section for the detail on DLT40 cadence}.
Given the magnitude limit of the program ($r\sim19\,\rm{mag}$) DLT40 is %quite sensitive to 
well suited to detect nearby kilonova event.% if they were occurring.
%A second 
An advantage of DLT40, is that we can directly use the light curve of AT~2017gfo obtained with the same instrumental set-up\citep{Valenti17}
%with the same filterless observations 
to get a direct limit for similar transients in the DLT40 program.
%This approach has several advantages.  First, the DLT40 search has a one day cadence \footnote{{\bf  see next section for the detail on DLT40 cadence}}.
%and focused target list of galaxies within $D\lesssim40\,\rm{Mpc}$ which, given the magnitude limit of the program ($r\sim19\,\rm{mag}$) makes it quite sensitive to nearby kilonova events if they were occurring. 

This letter is organized as follows. In Section~\ref{sec:search} we describe the DLT40 search and the survey operation during the O2 run. In Section~\ref{sec:rate} we will highlight the steps necessary to measure the rates, while in Section~\ref{sec:summary}, we will compare our results with previous rate estimates and discuss the future prospects on how the improve the rates with or independently from LIGO/Virgo next observing run.

\section{The DLT40 search}
\label{sec:search}
DLT40 is a one-day cadence\footnote{This is the goal however the reader should be aware that weather losses and technical problems may randomly affect our cadence of observations. As an example, during the first year of DLT40 14$\%$ nights were lost because of weather or technical problems.}  search for SNe, targeting galaxies in the nearby universe ($D\lesssim40\,\rm{Mpc}$)
 %with 
 and designed for the goal of discovering $\sim10$ SNe per year within one day of explosion \citep{2017arXiv171103940T}. 
% In reality DLT40 will discover roughly  $\sim20$ SNe per year but only half of them 
Running continuously since September 2016, DLT40 is observing $\sim$400-600 galaxies every night using a $0.41\,\rm{m}$ PROMPT telescope \citep{2005NCimC..28..767R} at the Cerro Tololo Inter-American Observatory (CTIO).

\begin{figure*}
\begin{center}
\includegraphics[width=0.75\textwidth]{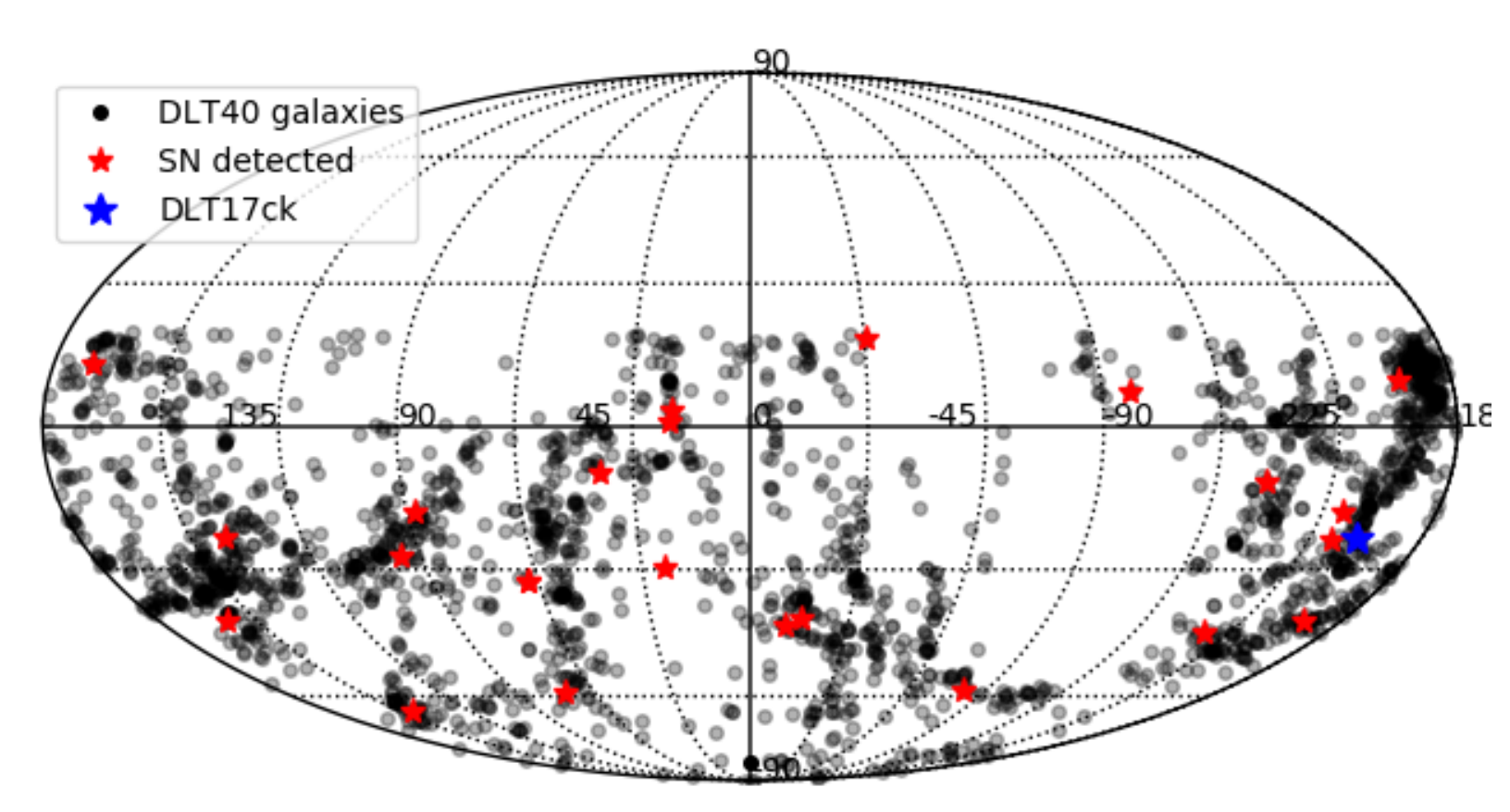}
\includegraphics[width=1\textwidth]{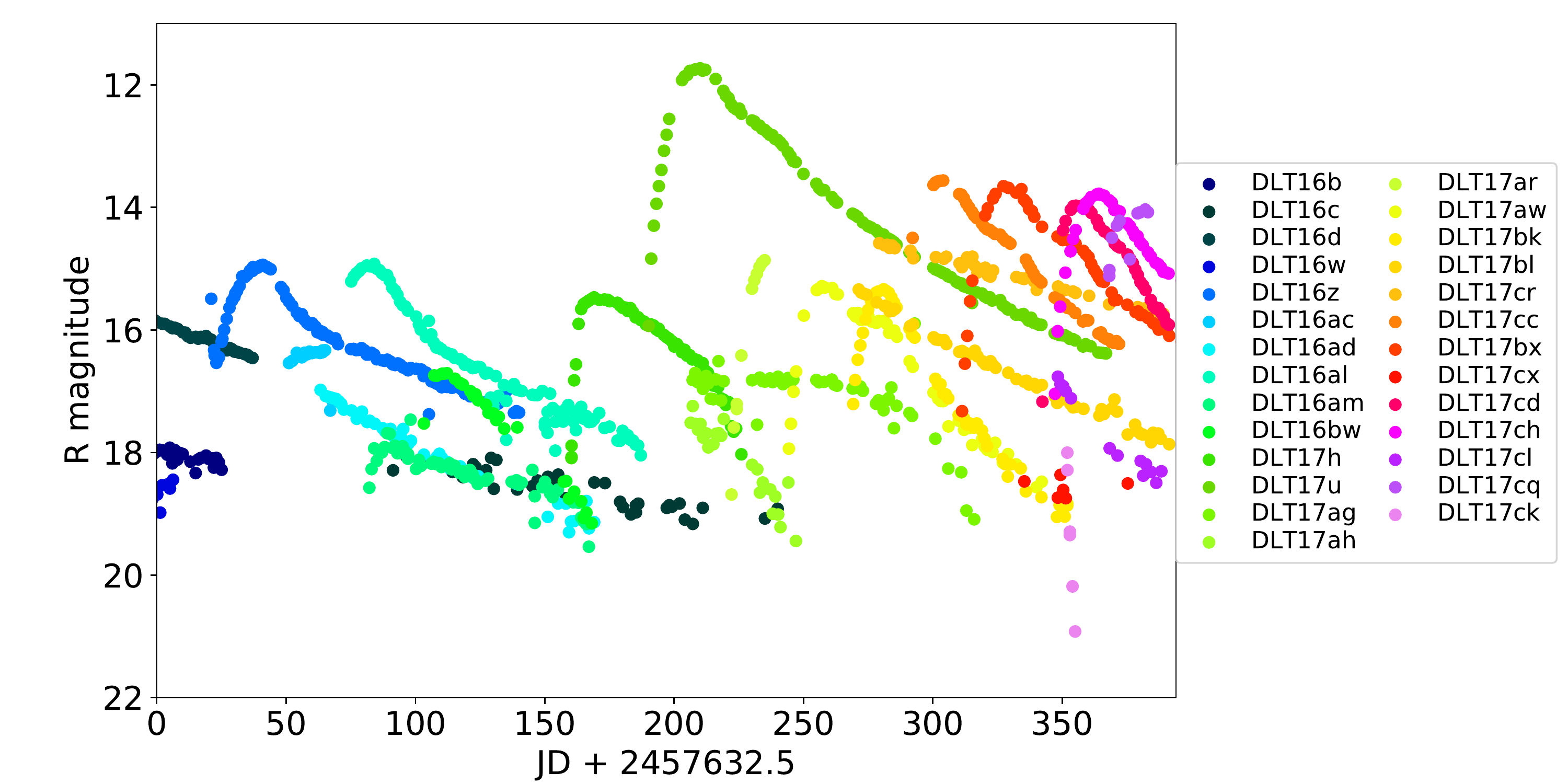}
\caption{Top panel: The DLT40 galaxy catalog (black points). The SNe discovered during the first year of DLT40 are also shown (red points) together with the kilonova DLT17ck (blue point). Lower panel: DLT40 light curves of all the SNe (and the kilonova) discovered during the first year of the search.}
\label{fig:dlt40}
\end{center}
\end{figure*}

With a field of view (FoV) of $10\times10\,\rm{arcmin^2}$, DLT40 is suitable to map nearby galaxies down to a limiting magnitude of $r\sim19\,\rm{mag}$ in $45\,\rm{s}$ unfiltered exposures.
%(with an open filter). 
The DLT40 catalog includes galaxies from the Gravitational Wave Galaxy Catalogue \citep[GWGC;][]{2011CQGra..28h5016W} at declination $<20\,\rm{degrees}$, with an absolute magnitude $M_{B}<-18\,\rm{mag}$, galactic extinction $A_V<0.5\,\rm{mag}$ and recessional velocity $v_r<3000\,\rm{km}\,\rm{s^{-1}}$ (corresponding to $D\lesssim40\,\rm{Mpc}$). An additional constraint is that the field of view should not include stars brighter than 8th mag. %These cuts leave 
These selections led to a sample of 2220 bright galaxies. %, all with distance estimates \citep[see discussion in][]{2011CQGra..28h5016W}, absolute magnitude and apparent diameter.  
In Figure \ref{fig:dlt40}, we show the sky distribution of galaxies in the DLT40 catalog. % over the sky. 
By comparing the integrated luminosity of DLT40's galaxy sample with the total integrated luminosity of the GWGC catalog, 
we estimated that the DLT40 catalog contains $\sim 60 \%$ of the total luminosity (and therefore roughly mass) of the GWGC catalog. Notice that the GWGC is supposed to be complete out to 40 Mpc \citep{2011CQGra..28h5016W}.
The DLT40 galaxy catalog is not biased with respect to morphological type of galaxy (Sheng et al. in preparation) but by construction it is biased against low luminosity galaxies (for example dwarf galaxies).

In order to find SNe within a day from explosion, the DLT40 data are processed in nearly real time. Pre-reduced images are transferred within $\sim1\,\rm{minute}$ to our dedicated server where new candidates are detected by image subtraction with respect to a template image \citep[using {\tt Hotpants}][]{Becker15}, so that new candidates are available for scanning by %the DLT40 group 
human operator within a few minutes after acquisition. Confirmation images of new SN candidates are taken within a few hours of their DLT40 discovery, often with the Las Cumbres Observatory telescope network \citep{Brown2013}.

Since the beginning of the search (September 2016), we have discovered 26 SNe, twelve of which were first announced by DLT40.  Seven were discovered within $48\,\rm{h}$ of explosion (see Table~\ref{tab:tab1} for a list of transients discovered by DLT40). The late discovery of the remaining  transients by DLT40 was due to poor weather conditions.

While searching for SNe, DLT40 also reacted to LIGO/Virgo triggers during the O2 observing run, prioritizing the galaxies from the DLT40 catalog within the LIGO/Virgo localization region for each trigger. Following the LIGO/Virgo trigger of GW170817 \citep{2017GCN..21509....1A, 2017GCN..21527....1A}, DLT40 independently discovered %(among other groups) 
and monitored the evolution of the kilonova AT~2017gfo/DLT17ck \citep{Valenti17}. Given the daily cadence of the search, DLT40 is well suited to discover similar fast kilonova-like transients. In the particular case of DLT17ck, we did have to  revise our target priority list because the GW localization placed it near the horizon at sunset in Chile.  During the course of the normal survey, however, any other kilonova-like transient would have been visible in the DLT40 search fields, given that DLT17ck itself was $\sim$1.5 magnitudes brighter than our typical detection limit, %even 
out to the border of our $D$\/$\sim$40 Mpc pointed search.

\section{Rate Measurement}
\label{sec:rate}
One approach to measuring the rate of an  astronomical transient makes use of the \emph{control time} concept \citep{1942ApJ....96...28Z, 1993A&A...268..472C, 1997A&A...322..431C}.
For each $i$-th galaxy, the control time ($ct_{i}$) is defined as the time during which a hypothetical transient is above the detection limit. It depends on the magnitude limit of each observation and the light curve of the transient. The total control time per unit luminosity of our search is computed by multiplying  the $ct_{i}$ control time by the luminosity of the $i$-th galaxy, and then summing over all of the galaxies in our sample:\footnote{The control time $ct_{i}$ depends also on the absolute magnitude of the transient (brighter transients will remain visible for a longer time above threshold).
%each component should in principle be obtained by computing the control time over the luminosity function of the transient ($ct_ij$). The control time for each galaxy is then  $ct_i =  \sum_{i=1}^{n}ct_ij * f_j$  where $f_j$ is the fractional number of transients with a magnitude in the $j$-th luminosity function bin. However,  
To account for the transient luminosity function, we use a monte carlo approach, simulating a number of transients for each galaxy following %the transient luminosity function. Therefore we do not need to add this term in our computation.
an adopt distribution}

\begin{eqnarray} 
ct=\sum_{i=1}^{n}L_i*ct_i
\nonumber
\end{eqnarray}

%Each component to the total control time, should indeed be weighted by the fractional contribution to the luminosity function of the transient.}
The ratio between the number of transients detected and the sum of the control times for all galaxies observed gives immediately the rate as:
\begin{eqnarray} 
r=\frac{N}{ct}
\nonumber
\end{eqnarray}

In order to measure the control time, the first step is evaluate the transient detection efficiency for each image or, in other words, to measure the apparent magnitude limit for transient detection. In order to do that, we performed artificial star experiments for a subset of frames, implanting stars with different magnitudes using the proper point-spread functions (PSFs), and registering the fraction of artificial stars automatically identified by our pipeline on the difference images. 

\begin{figure*}
\begin{center}
\includegraphics[width=0.45\textwidth]{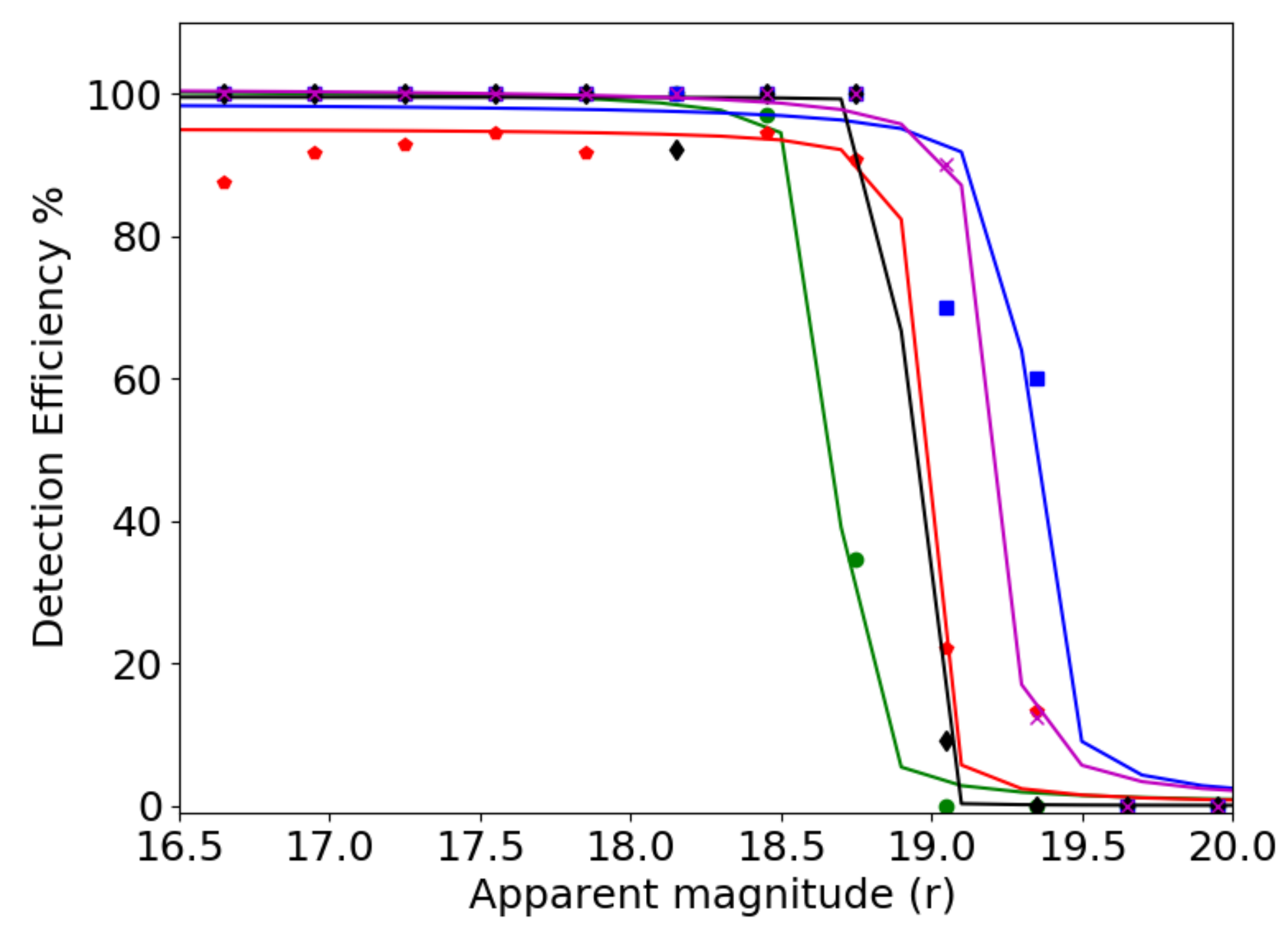}
\includegraphics[width=0.45\textwidth]{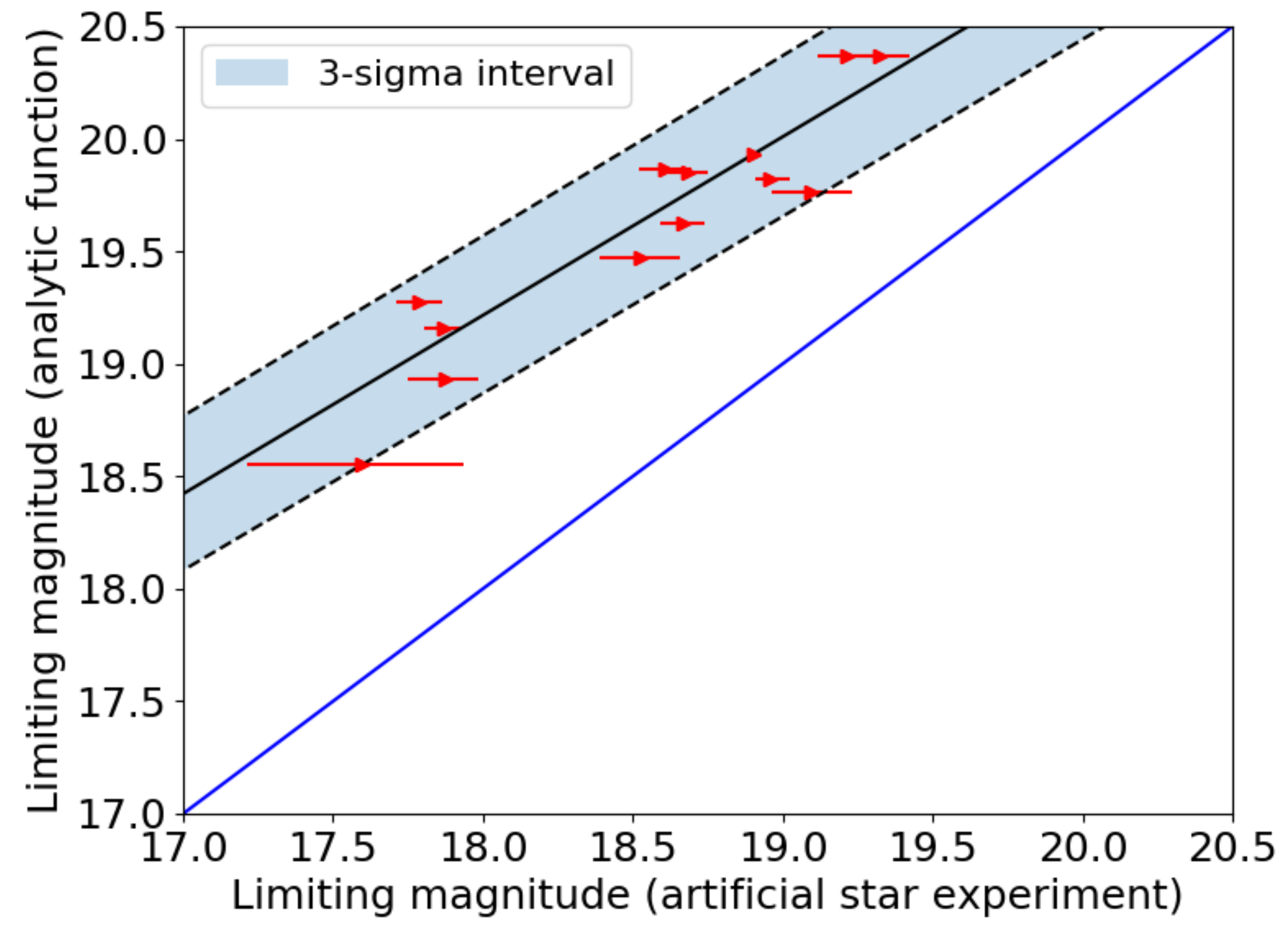}
\caption{Left panel: Transient detection efficiency as a function of apparent magnitude for 5 DLT40 fields. The lines are the best fit to the curve. The limiting magnitude is chosen at 50$\%$ efficiency. Right panel: We compare the limiting magnitude computed for each image using its zeropoint and an analytical function with the limiting magnitude computed with artificial star experiments on difference images. This linear relation has been used to scale the  limiting magnitude computed for each frame (given its zeropoint) to a more realistic limiting magnitude estimate for SN and/or kilonova detection.}
\label{fig:efficiency}
\end{center}
\end{figure*}

Hereafter we will adopt the magnitude corresponding to a $50\%$ detection efficiency as the limiting magnitude for the DLT40 survey, while we use the $16\%$ and $84\%$ detection efficiencies as lower/upper limits to determine its uncertainty. We found that the magnitude limit of our search is in the range $M_{r}\sim18-20\,\rm{mag}$ (see left panel of Figure~\ref{fig:efficiency}) depending on weather and seeing conditions of the specific observation.  Since artificial star experiments are time consuming, instead of repeating the simulation for all of the $\sim120,000$ frames observed so far, we exploited a linear relation between the limiting magnitude  for transient detection computed through artificial star experiments and the limiting magnitude for stellar source detection computed for each target frame (i.e. not the difference image). The latter was derived through an analytic equation  using information on the noise and photometric calibration for each image. The comparisons between the two limiting magnitudes is shown in 
the right panel of Figure \ref{fig:efficiency}. In general, the limiting magnitude computed with the analytic function on the target image (y axes) is $\sim$1 magnitude deeper than the limit magnitude from artificial stars experiment (x axes). This is expected since the difference imaging technique effectively adds the template image noise to that of the target image.%, roughly in quadrature.}

%(see right panel of Figure \ref{fig:efficiency}).  The $\sim$1 magnitude difference between the two values is entirely due to the difference imaging technique employed, which effectively adds the template image noise to that of the target image, roughly in quadrature.

The second ingredient to measure the control time is simulation of kilonova light curves in the time window each galaxy was observed. The time that the transient is above our detection limit contributes to the control time. The observed light curve of AT~2017gfo/DLT17ck was used as a reference, scaled to the distance of each galaxy with an explosion epoch randomly distributed in the observed time window.
%with a $0.5\,\rm{day}$ sampling interval in each galaxy, with an explosion epoch randomly distributed in the observed time window. 
We took into account that kilonovae may have a range of absolute magnitudes, and that they may experience a variety of host galaxy extinction due to dust.  For the range in kilonova magnitudes, we varied the absolute magnitude of the kilonova using a Gaussian distribution centered on the absolute magnitude of 
AT~2017gfo/DLT17ck and a sigma of 0.5 magnitudes (e.g. 95\% of simulated light curves have an absolute magnitude within  $\pm1\,\rm{mag}$ of AT~2017gfo/DLT17ck).
For the extinction distribution, we notice that the host environment of neutron stars mergers is often compared to the host environment of SNe Ia since both types of systems are found in early-type and star-forming galaxies \citep{2013ApJ...769...56F}. For this reason, we adopted for the extinction distribution $P(A_V)=e^{-A_V/\tau_V}$, with $\tau_V=0.334\pm0.088\,\rm{mag}$\citep{2009ApJS..185...32K}, which we label `SN Ia extinction' scenario.  
We also computed the control time using either no extinction (low extinction scenario) or an extinction distribution with a $\tau$ value inflated by a factor 2 (high extinction scenario).
We want to stress that, giving that DLT17ck is the first clear case of a kilonova, any choice of absolute magnitude range and reddening law is  somehow 
arbitrary and those quantities will be better constrained when a larger number of kilonovae was discovered.

In summary, for each galaxy, we have simulated 20,000 light curves randomly distributed in the 13 months of the search, with a range of absolute magnitudes and reddening. If at any epoch of observation, the simulated light curve was brighter than our detection limit, the simulated transient would have been detected. The fraction of detected simulated transients, multiplied by the time window each galaxy was observed, gives the control time. The uncertainty on the detection limits (right panel of Figure~\ref{fig:efficiency}), are reported as systematic errors, while the three extinction distributions used (low, similar to SNe Ia and  high extinction) are reported separately.
%{\bf We have further investigated that our choice of magnitude range have only a small effect on the control time.}
% The lower and higher control time values, varying absolute magnitude and reddening, are used as lower and upper limits for the error.  Nonetheless, Since we observe only galaxies within $40\,\rm{Mpc}$ and DLT17ck was $1.7\,\rm{mag}$ brighter than our limiting magnitude, 
%Given the fast evolving light curve of the kilonova AT~2017gfo/DLT17ck, our control time per galaxy is on average $80.\pm 5,\rm{days}$.

During the 13 months of the search, the average number of observed frames per galaxy was 64, while the average control time per galaxy was 80 days. This means that any fast evolving transient like AT~2017gfo/DLT17ck would likely not be detected a second time if the survey cadence was 2 days or larger. Our strategy of triggering a confirmation image for each new target within a few hours of first detection well fits the need for these fast transients. Excluding AT~2017gfo/DLT17ck, which was discovered only thanks to the LIGO/Virgo trigger, no other transient with a similar fast evolution was detected.
%As the discovery of AT~2017gfo/DLT17ck by DLT40 was possible only thanks to the LIGO/Virgo trigger, we do not consider this detection in our rate, and since no other transient with a similar fast evolution was detected, 
We infer a limit to the rate of kilonovae of $<0.47\,_{-0.03}^{+0.04}\,\rm{SNuB}$\footnote{$\rm{SNuB}=1\,\rm{SN}$ per $100\,\rm{yr}$ 
per $10^{10}L_{B_{\odot}}$}(low extinction), $<0.50\,_{-0.04}^{+0.05}\rm{SNuB}$ (SNe Ia extinction) and $<0.55\,_{-0.05}^{+0.07}\rm{SNuB}$ (high extinction), where the rate has been normalized to the galaxy integrated luminosity. This translates to a limit in our Galaxy of $<0.94\,_{-0.37}^{+0.38}$ (low extinction),  $<1.00\,_{-0.36}^{+0.43}$ (SN Ia extinction),  $<1.10\,_{-0.40}^{+0.51}$ (high extinction) 
per 100 years. These limits and the systematic error are reported in Table~\ref{tab:rate}. As a cross check, we have also computed from DLT40 the SN rates for SNe~Ia, Ibc and II  that will be reported in a dedicated paper (Yang et al in preparation). We stress that our SN rates estimates  are consistent with previous measurements \citep{1993A&A...268..472C,1997A&A...322..431C, 2011MNRAS.412.1419L}, despite the poor statisitcs a few simplifications in the calculation of the control time.
%In the same table, we also report the rates of SNe~Ia, Ibc and II measured in our survey. More details on the SN rates derived from the DLT40 survey will be reported in a companion paper (Yang et al in preparation), but here we stress that our SN rates, allowing for the small statistics, are consistent with previous measurements \citep{1993A&A...268..472C,1997A&A...322..431C, 2011MNRAS.412.1419L} suggesting that, despite a few simplifications in the calculation of the control time, our rates and limit are reliable. 

\section{Summary and Future Prospects}
\label{sec:summary}

\begin{figure*}
\begin{center}
\includegraphics[width=0.95\textwidth]{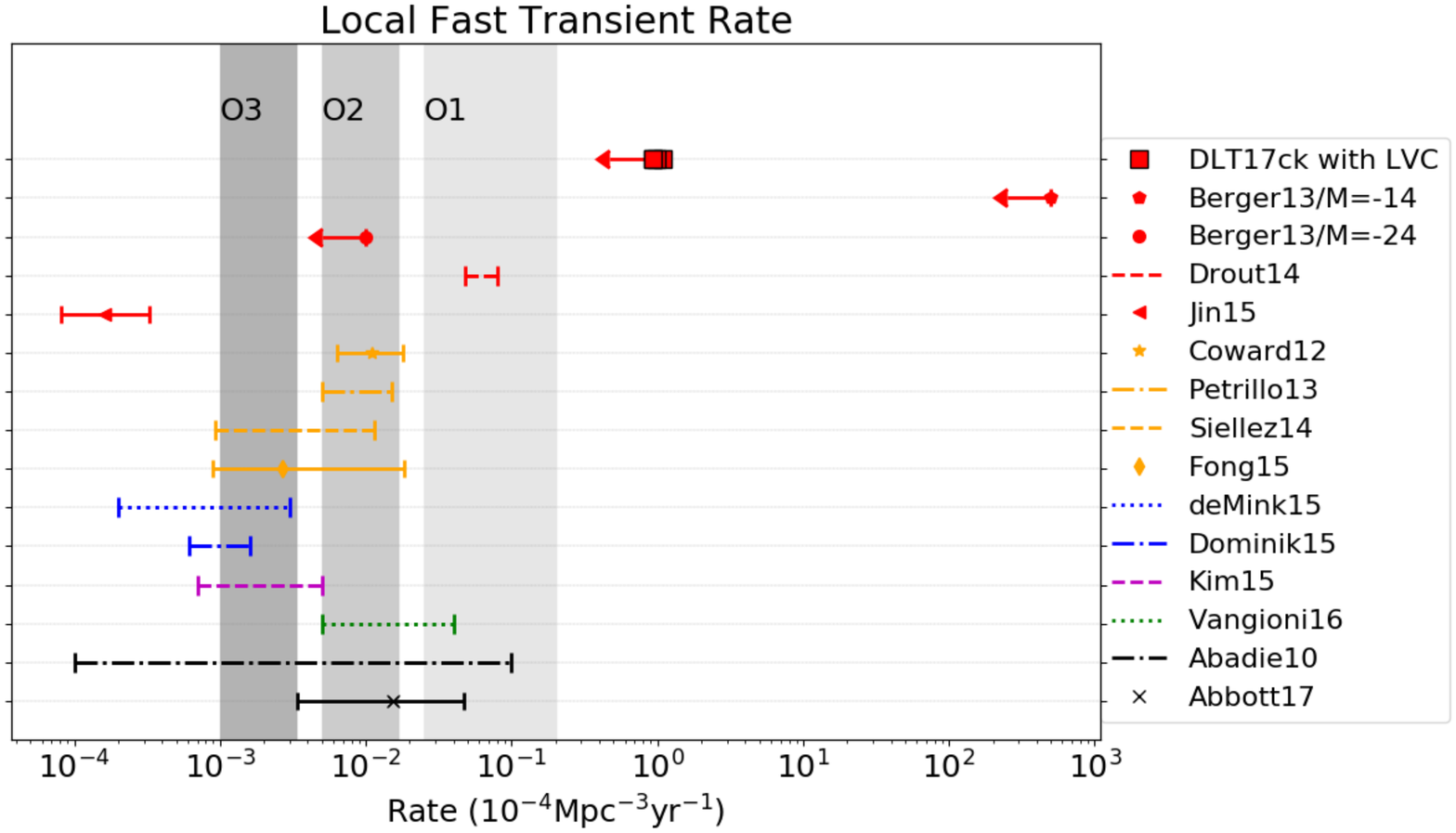}
\caption{DLT40 limit on the kilonova rate (all three reddening scenarios) compared with the rate of sGRB \citep[orange symbols, ][]{2012MNRAS.425.2668C, 2013ApJ...767..140P, 2014MNRAS.437..649S, 2015ApJ...815..102F}, the rates of BNS merger from stellar evolution \citep[blue lines, ][]{2015ApJ...814...58D, 2015ApJ...806..263D}, cosmic nucleosynthesis \citep[green line, ][]{2016MNRAS.455...17V}, galactic pulsar population \citep[magenta line, ][]{2015MNRAS.448..928K}, gravitational waves \citep[black lines, ][]{2010CQGra..27q3001A, 2017PRL119161101}  and fast optical transients \citep[red symbols, ][]{2015ApJ...811L..22J,2014ApJ...794...23D, 2013ApJ...779...18B}. }
\label{fig:rate}
\end{center}
\end{figure*}

In this paper, we have used the observed light curve of a kilonova to constrain the rate of BNS mergers to less than $0.47\,_{-0.03}^{+0.04}\,\rm{SNuB}$ (low extinction), $0.50\,_{-0.04}^{+0.05}\rm{SNuB}$ (SNe Ia extinction) and $0.55\,_{-0.05}^{+0.07}\rm{SNuB}$ (high extinction).
Since some published measurements of the BNS coalescence rates are expressed in units of co-moving volume, we convert SNu rates to volumetric rates similarly to \cite{Botticella12}, %We assume that the BNS coalescence rate is proportional to the $B-$band luminosity of the host galaxies, and converted the rates in units of SNuB to volumetric rates 
that is multiplying the SNuB rate by the galaxy $B-$band luminosity density reported in \cite{Kopparapu2008} $(1.98\pm0.16)\times10^{-2}\times10^{10}\,L^{B}_{\odot}\,Mpc^3$. The kilonova volumetric rate upper limit is 
$0.93\,_{-0.18}^{+0.16}\,10^{-4}\,\rm{Mpc^{-3}}\,\rm{yr^{-1}}$,  $0.99\,_{-0.15}^{+0.19}\,10^{-4}\,\rm{Mpc^{-3}}\,\rm{yr^{-1}}$ or $1.09\,_{-0.18}^{+0.28}\,10^{-4}\,\rm{Mpc^{-3}}\,\rm{yr^{-1}}$ (depending on the extinction law used) and is compared with previous measurements in Figure \ref{fig:rate}. Our rate is one order of magnitude higher than the BNS merger rate limit obtained by LIGO/Virgo during the observing run O1 \citep{2010CQGra..27q3001A} and two order of magnitude higher than the optimistic rates of short Gamma-ray bursts \citep{2012MNRAS.425.2668C, 2013ApJ...767..140P}.

We can also investigate how long it would on average take for our search to discover (independently from LIGO/Virgo) a kilonova.  During the LIGO O2 run ($\sim1\,\rm{yr}$), $117\,\rm{d}$ of simultaneous LIGO-detector observing time has been used to discover one BNS coalescence \citep{2017PRL119161101}, { which means there are 1/(117/365)=3.12 BNS sources in the LIGO searching volumn}, while our control time for kilonovae in the same period (monitoring galaxies within $40\,\rm{Mpc}$) is $0.22\,\rm{yr}$ (on average 80 days per year per galaxy). Comparing the total luminosity of the DLT40 galaxy sample and the total luminosity of the GWGC catalog, within 40 Mpc, gives $60\%$ of the GWGC catalog sample monitored by the DLT40 survey.  In order to independently discover a kilonova we would need to run the DLT40 for 3.12 / (control time * completeness ) $\times$ the volume ratio of the two surveys aLIGO/aVirgo and DLT40.
During the O2 run, aLIGO/aVirgo were sensitive up to a volume of $78.5\,\rm{Mpc}$ \citep{2016ApJ...832L..21A} and taking into account the different volumes of the two surveys $(78.5/40)^{3}$, we would need to run DLT40 for $\sim 18.4\,\rm{years}$ in order to independently discover a kilonova. This explains why historical optical searches \citep[like the Lick SN search;][]{2011MNRAS.412.1419L} never detected a kilonova. 

Given that it is quite unlikely to independently discover a kilonova with a search like DLT40, we may at least understand what a DLT40-like search may be able to do during the O3 aLIGO/aVirgo run.  During O3, LIGO will be able to detect BNS coalescences out to $150\,\rm{Mpc}$, while Virgo should be sensitive out to $65-85$ $\,\rm{Mpc}$ \citep{2016LRR....19....1A}. If all kilonovae would be as bright as DLT17ck, with the current DLT40 observing strategy, we could detect kilonovae within a distance of $70\,\rm{Mpc}$.  In order to cover the full Virgo volume ($85\,\rm{Mpc}$), we would need to go $\sim0.4\,\rm{mag}$ deeper (to a limiting magnitude $\sim19.4\,\rm{mag}$), hence increasing the exposure time by a factor of 2.2 (100 seconds per exposure, instead of the current 45 seconds).  As DLT40 currently observes 400-600 galaxies per night with $45\,\rm{s}$ exposures, increasing the exposure time to 100 seconds would still allow us to observe $\sim230$ galaxies during a single night. Randomly selecting galaxies within $85\,\rm{Mpc}$ from the GLADE\footnote{http://aquarius.elte.hu/glade/} catalog in typical aLIGO/aVirgo regions (30 sq degrees) the average number of galaxies is $\sim230$ --  almost exactly the same number of galaxies observable by DLT40 each night with an exposure time of 100 seconds. Here we neglect that the GLADE catalog is only $\sim 85-90\%$ complete in the integrated luminosity up to $85\,\rm{Mpc}$ (GLADE catalog).

Therefore, within $85\,\rm{Mpc}$, small telescopes can still play a useful role (unless DLT17ck turns out to be a particularly bright kilonova), but the incompleteness of the available catalogs, especially for faint galaxies, may suggest that a wide-field of view strategy to directly tile the full aLIGO/aVirgo localization region may be preferred to avoid possible biases in sampling of the stellar population. In this respect the association of GRBs \citep{2009ApJ...691..182S} and SLSN \citep{2016ApJ...830...13P} with dwarf galaxies is a lesson learned.

\acknowledgments
Research by DJS and L.T. is supported by NSF grants AST-1412504 and AST-1517649. The DLT40 web-pages structure were developed  at the Aspen Center for Physics, which is supported by National Science Foundation grant PHY-1066293 AC acknowledges support from the NSF award $\#$1455090 "CAREER: Radio and gravitational-wave emission from the largest explosions since the Big Bang". The work of SY was supported by the China Scholarship Council(NO. 201506040044).

\bibliographystyle{apj}
%\bibliography{references_test}

\clearpage

\begin{table}
\caption{Summary table of the supernovae detected with DLT40. Their light curves are shown in Figure \ref{fig:dlt40}. Supernovae detected in background galaxies are marked as BKG.}
\centering
\normalsize
\label{snsum}
\begin{tabular}{lllllllll}
\hline
RA & DEC & DLT NAME & TNS NAME & SN TYPE & HOST GALAXY \\ %& OBS WINDOW(JD)\\
\hline
\hline
%204.52 & -17.85 & DLT16a & 2016C & SN IIp & NGC5247 & 2457559.71--2457612.56\\
%\hline
278.63 & -58.53 & DLT16b & 2016bmi & SN IIp & IC4721 \\%& 2457546.90--2457657.56\\
\hline
170.08 & 12.98 & DLT16c & 2016cok & SN IIp & NGC3627 \\%& 2457547.59--2457872.50\\
\hline
329.77 & 18.19 & DLT16d & 2016coi & SN Ic & UGC11868 \\%& 2457555.89--2457676.51\\
\hline
%203.66 & -23.68 & DLT16f & 2016eiy & SN Ia & ESO509-064 & 2457598.61--2457624.54\\
%\hline
328.62 & -57.66 & DLT16w & 2016fjp & SN Ia & BKG \\%& 2457626.76--2457639.74\\
\hline
23.56 & -29.44 & DLT16z & 2016gkg & SN IIb & NGC0613 \\%& 2457653.54--2457773.56\\
\hline
20.55 & 0.95 & DLT16ac & 2016hgm & SN II & NGC0493 \\%& 2457682.83--2457701.55\\
\hline
140.87 & -23.17 & DLT16ad & 2016gwl & SN Ia & NGC2865 \\%& 2457687.84--2457801.58\\
\hline
63.02 & -32.86 & DLT16al & 2016iae & SN Ic & NGC1532 \\%& 2457707.66--2457819.62\\
\hline
63.03 & -32.85 & DLT16am & 2016ija & SN II & NGC1532 \\%& 2457713.69--2457799.54\\
\hline
114.29 & -52.32 & DLT16bw & 2016iyd & SN II & BKG  \\%& 2457735.77--2457801.58\\
\hline
159.32& -41.62 & DLT17h & 2017ahn & SN II & NGC3318 \\%& 2457792.79--2457862.59\\
\hline
218.14 & -44.13 & DLT17u & 2017cbv & SN Ia & NGC5643 \\%& 2457822.64--2457999.49\\
\hline
193.46 & 9.70 & DLT17ag & 2017cjb & SN II & NGC4779 \\%& 2457838.87--2457895.47\\
\hline
200.52 & -13.14 & DLT17ah & 2017ckg & SN II & BKG \\%& 2457838.88--2457879.59\\
\hline
144.15 & -63.95 & DLT17ar & 2017cyy & SN Ia & ESO091-015 \\%& 2457854.68--2457867.49\\
\hline
263.11 & 7.06 & DLT17aw & 2017drh & SN Ia & NGC6384 \\%& 2457874.87--2457984.52\\
\hline
192.15 & -41.32 & DLT17bk & 2017ejb & SN Ia & NGC4696 \\%& 2457901.72--2457984.48\\
\hline
349.06 & -42.57 & DLT17bl & 2017bzc & SN Ia & NGC7552 \\%& 2457903.86--2457998.62\\
\hline
344.32 & -41.02 & DLT17cr & 2017bzb & SN II & NGC7424 \\%& 2457906.88--2458014.80\\
\hline
227.31 & -11.33 & DLT17cc & 2017erp & SN Ia & NGC5861 \\%& 2457923.53--2457999.48\\
\hline
20.06 & 3.40     & DLT17bx & 2017fgc & SN Ia & NGC0474 \\%& 2457943.79--2457998.67\\
\hline
114.11  & -69.55 & DLT17cx & 2016jbu & SN IIn & NGC2442 \\
\hline
95.39 & -27.21 & DLT17cd & 2017fzw & SN Ia & NGC2217 \\%& 2457974.86--2457998.80\\
\hline
71.46 & -59.25 & DLT17ch & 2017gax & SN Ibc & NGC1672 \\%& 2457979.74--2457998.70\\
\hline
88.27 & -17.87 & DLT17cl & 2017gbb & SN Ia & IC0438 \\%& 2457980.84--2457985.89\\
\hline
38.88 & -9.35 & DLT17cq & 2017gmr & SN II & NGC0988 \\%& 2458000.70--2458016.74\\
\hline
197.45 & -23.38 & DLT17ck & 2017gfo & kilonova & NGC4993 \\%& 2457984.49--2457987.55\\
\hline
\end{tabular}\\
\end{table}

\begin{table*}
\newcommand{\tabincell}[2]{\begin{tabular}{@{}#1@{}}#2\end{tabular}}
\centering
\begin{minipage}{150mm}
\caption{DLT40 rate estimation results}
\label{tab:rate}
\scriptsize
\begin{tabular}{| l | l | l | l | l | l |} 
\hline
Type & 
\tabincell{l}{extinction\\(mag)}  &
\tabincell{l}{control time\\(days)} &
\tabincell{l}{lums rate$^{a}$\\(SNuB)} &
\tabincell{l}{vol rate$^{b}$\\($10^{-4} Mpc^{-3} yr^{-1}$)}& 
\tabincell{l}{Milky Way rate$^{c}$\\($(100yr)^{-1}$)}\\
\hline
no reddening & $P(A_V) = 0$ & $79.67^{+4.51}_{-5.53}$ & $<0.47^{+0.04}_{-0.03}$ & $<0.93^{+0.16}_{-0.18}$ & $<0.94^{+0.38}_{-0.37}$ \\
Ia reddening & $P(A_V)=e^{-A_V/0.334}$ & $75.07^{+5.35}_{-6.56}$ & $<0.50^{+0.05}_{-0.04}$  & $<0.99^{+0.19}_{-0.15}$ & $<1.00^{+0.43}_{-0.36}$\\
high reddening & $P(A_V)=2 \times e^{-A_V/0.334}$ & $69.46^{+6.15}_{-7.28}$ & $<0.55^{+0.07}_{-0.05}$ & $<1.09^{+0.24}_{-0.18}$ & $<1.10^{+0.51}_{-0.40}$\\
\hline
\end{tabular}
\label{tab:tab1}\\
(a) : DLT40 only detected DLT17ck because of the LIGO detection and subsequent localization, therefore it is not considered in our rate calculations, which we report here as 95\% confidence level Poisson single-sided upper limits, given zero events \citep{1986ApJ...303..336G}.\\
(b) : We converted the rates in units of SNuB to volumetric rates with luminosity density: $(1.98\pm0.16)\times10^{-2}\times10^{10}  L^{B}_{\odot} Mpc^3$ \citep{Kopparapu2008}.  \\
(c) : The total B-band luminosity of the MW is quite uncertain; we adopt $(2.0\pm0.6)\times10^{10}  L^{B}_{\odot}$ \citep{1987A&A...173...59V}.\\
\end{minipage}
\end{table*}

\end{document}